\documentclass[twocolumn,english,aps,prb,twocolum,superscriptaddress,amsmath,amssymb,floatfix]{revtex4-2}

\pdfoutput=1
\usepackage[utf8]{inputenc}
\usepackage{amsfonts}
\usepackage{amsmath}
\usepackage{amssymb}
\usepackage[colorlinks=true,citecolor=blue,linkcolor=blue,breaklinks=true]{hyperref}
\usepackage{graphicx}

\usepackage{soul}
\usepackage{url}
\usepackage{dcolumn}
\usepackage{bm}
\usepackage{stmaryrd}
\usepackage[final]{changes}
\usepackage{multirow}
\usepackage{rotating}
\usepackage{titlesec}
\usepackage{tabularx}
\usepackage{textcomp}
\usepackage[flushleft]{threeparttable}
\usepackage{bibunits}

\usepackage{titlesec}
\renewcommand{\thesection}{\arabic{section}}
\titleformat{\section}{\normalfont\large\bfseries}{\thesection.}{1em}{}

\newcommand{\ii}{\mathrm{i}}
\newcommand{\bs}{\boldsymbol}

\usepackage{natbib}
\begin{document}

\begin{bibunit}[plain]

\title{Realization of tilted Dirac-like microwave cone in superconducting circuit lattices}

\author{Amir Youssefi}\thanks{These authors contributed equally.}
\affiliation{Laboratory of Photonics and Quantum Measurement, Swiss Federal Institute of Technology Lausanne (EPFL), Lausanne, Switzerland}
\affiliation{EDWATEC SA, Lausanne, Switzerland}
\author{Ahmad Motavassel}\thanks{These authors contributed equally.}
\affiliation{EDWATEC SA, Lausanne, Switzerland}
\author{Shingo Kono}
\affiliation{Laboratory of Photonics and Quantum Measurement, Swiss Federal Institute of Technology Lausanne (EPFL), Lausanne, Switzerland}
\affiliation{Center for Quantum Science and Engineering, EPFL, Lausanne, Switzerland}
\author{Seyed Akbar Jafari}
\email[]{akbar.jafari@rwth-aachen.de}
\affiliation{II. Physikalisches Institut C, RWTH Aachen University, 52074, Aachen, Germany}
\author{Tobias J. Kippenebrg}
\affiliation{Laboratory of Photonics and Quantum Measurement, Swiss Federal Institute of Technology Lausanne (EPFL), Lausanne, Switzerland}
\affiliation{Center for Quantum Science and Engineering, EPFL, Lausanne, Switzerland}

\begin{abstract}
Dirac-like band crossings are paradigms in condensed matter systems to emulate high-energy physics phenomena. They are associated with two aspects: gap and tilting. The ability to design sign-changing gap gives rise to band topology, whereas the tilting of band crossings which is a gateway for large gravity-like effects remains uncharted. In this work, we introduce an experimental platform to realize tilted Dirac-like microwave cone in large-scale superconducting circuit lattices. The direction and magnitude of the tilt can be controlled by engineering the axially preferred second neighbor coupling. We demonstrate three lattices with $731$-site LC resonator featuring tilt values of up to $59\%$ of relative difference in the opposite-direction group velocities. This is obtained by reconstructing the density of states (DOS) of measured microwave resonance frequencies. Harnessing the tilt of Dirac-like band crossings lays the foundation for weaving the fabric of an emergent solid-state spacetime. 
\end{abstract}

\maketitle


\section{Introduction}
\begin{figure}[t]
	\centering
	\includegraphics[width = \columnwidth]{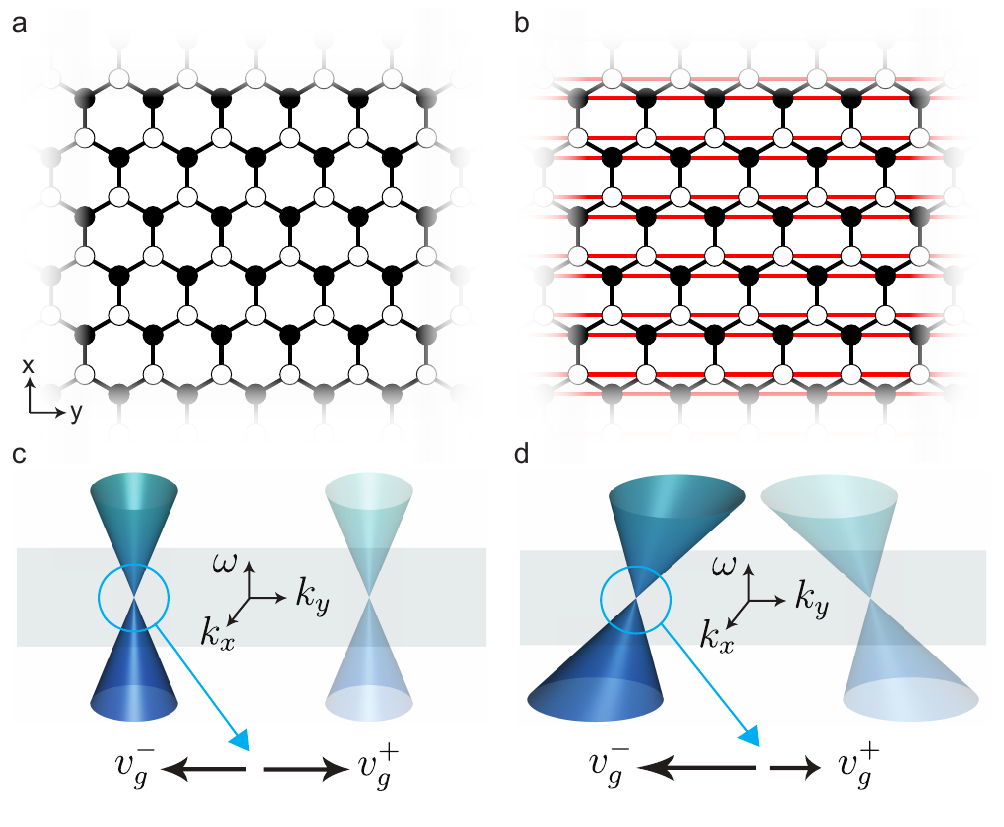}
	\caption{ \textbf{Lattice structure and resulting microwave cones} \textbf{a}, simple honeycomb lattice with symmetric first neighbor coupling. \textbf{b}, A second neighbor coupling is added along a preferred direction shown by the red lines in the $y$ direction. \textbf{c-d}, Dirac-like microwave cone for a,b lattices, respectively. The group velocities for the left cone at $K$ point of the Brillouin zone in the $y$ direction are  $v_g^\pm$. Introducing non-symmetric second neighbor coupling results in the tilting of this Dirac-like microwave cone in $+y$ direction, giving rise to different group velocities $v_g^\pm$ . The other cone at K' (light blue) is tilted oppositely.}
    \label{fig:1}
\end{figure}

Propagation of photons in flat spacetime is characterized by the light cones~\cite{Rindler2006} that are associated with the energy-momentum or dispersion relation $\varepsilon(\mathbf{p})=c|\mathbf{p}|$~\cite{MoldingLight} where $c$ is the speed of light. According to Einstein's general relativity, in the presence of a gravity source, the light cones are tilted towards it~\cite{Ryder2009}, implying that the propagation velocity towards ($v_g^-$) and away ($v_g^+$) from the gravitational source will not be equivalent anymore. The amount and direction of the tilting, quantified by a vector $\bs\zeta$ whose magnitude $\zeta=2|v_g^--v_g^+|/(v_g^-+v_g^+)$ is an indicator of the presence of gravity (see Fig~\ref{fig:1}d) which defines the structure of the underlying spacetime metric. However, effects of gravity on light are generally diminished by small gravity constant $G$ and large speed of light $c$~\cite{Roy2019}. Synthesis of a strongly curved spacetime can be a useful framework to investigate fundamental questions such as the quantum entanglement generated by spacetime curvature~\cite{Aspelmeyer2023}, behavior of waves~\cite{Friedlander1975} and quantum fields in strongly curved geometry~\cite{Birrell1982}, measurements of the stronger form of various properties of the curved spacetimes such as gravitational red-shift~\cite{Pound1960} and properties of blackholes~\cite{Carroll2019}. Strongly curved synthetic spacetime can even be envisaged to employ frame-dragging properties such as Lens-Thirring~\cite{Ryder2009} effect in spintronics applications.

Certain periodic solid-state structures show cone-shaped dispersion relations. Prominent examples are Dirac and Weyl materials: In quantum materials such as graphene~\cite{Katsnelson2020,Aoki2014} and its 3D counterparts collectively known as Dirac~\cite{Armitage2018} or Weyl materials~\cite{YanFelser2017}, the energy-momentum dispersion relation in the continuum limit is given by the solid-state analog of the dispersion of light $\varepsilon(\mathbf{p})=v|\mathbf{p}|$, except that $v$ is Fermi velocity which is usually $2-3$ orders of magnitude smaller than the speed of light $c$. The upright Dirac cone is obtained when the long-wavelength limit is isotropic. However, in the presence of additional anisotropic hopping, the cone-shaped crossing can be tilted.  Therefore an emergent metric associated with tilting $\bs\zeta$ (see SI) is a universal description of tilted bands irrespective of whether the tilting is caused by gravity or solid-state engineering. The advantage of solid-state platforms over the actual gravity-based spacetime is that, first the tilting parameter can be as large as tens of percents~\cite{Jafari2023}. Secondly, the $\bs\zeta$ can be tuned by the appropriate design or external fields~\cite{Ogawa2016}. 

In order to obtain a tilted cone from the upright cone in a honeycomb structure, one has to break the rotational symmetry by equally hopping-coupling the second neighbors on both sub-lattices in a preferred direction\cite{Yekta2023,mota2021circuit}.  Quantum materials with such a tilted band crossing include organic materials~\cite{Kajita2014}, hydrogenated graphene~\cite{Lu2016}, the so-called $8Pmmn$ borophene~\cite{Zhou2014,LopezBezanilla2016}, 2D quantum wells~\cite{Tao2018} and surface of transition metals~\cite{Miyamoto2012,Mirhosseini2013,Varykhalov2017}. The tilt of the Dirac cone is responsible for many unusual properties~\cite{Suzumura2015,Yang2018,Wild2022,Islam2017,Kundu2020,Trescher2015,Sengupta2018}. The most striking of such effects would be an electron transport induced by the time-gradient of temperature, $\partial_tT$~\cite{Moradpouri2023} or a tilt-induced vortical anomaly~\cite{Rostamzadeh2023} or violation of fundamental bounds of nature~\cite{HolographicMoradpouri}. Another family would be Weyl semimetals with $|\bs{\zeta}|>1$  known as type-II Weyl semi-metals that have been considered to represent the interior of blackholes~\cite{Volovik2016,Volovik2017,Zubkov2018,Huang2018,Hashimoto2020}.  The tilting parameter is a property of a given material and can only be incrementally changed by perturbations such as strain~\cite{Manes2013}. Other proposals on selective replacements of boron atoms in borophene compounds require atomic scale precision in substitutional doping~\cite{Yekta2023}, and therefore remain challenging.  Tuning the tilting of the Dirac cone by electromagnetic fields~\cite{Farajollahpour2019} leads to undesirable complications, since the mass of the Dirac particle would also depend on the applied perpendicular electric field. The spin-orbit coupled Dirac cones on the surface of topological insulators offer a promising platform for the engineering of the tilt of the Dirac cone by external magnetic influence~\cite{Ogawa2016,Jafari2023}. Therefore, a crucial step towards the synthesis of arbitrary spacetime structure in solid-state systems is reliable engineering of the tilt of the band crossing at will. 

Superconducting circuit lattices have been an attractive platform to implement and study emerging phenomena in condensed matter physics~\cite{carusotto2020photonic,sheremet2023waveguide} and quantum simulation~\cite{schmidt2013circuit,altman2021quantum}, because of their design flexibility, low microwave losses, operating at the quantum ground-state, and integrability with superconducting qubits. Particularly, one- or two-dimensional superconducting networks have been used to realize topological waveguides~\cite{kim2021quantum}, quantum phase transitions~\cite{fazio2001quantum}, atom-photon bound states~\cite{scigliuzzo2022controlling}, strained graphene models~\cite{youssefi2021superconducting}, and hyperbolic lattices~\cite{kollar2019hyperbolic}. However, those platforms are compounded by scaling limitations due to the large lattice unit cell footprint area, and the frequency and coupling disorder. Superconducting co-planar waveguide (CPW) resonators~\cite{underwood2012low,underwood2016imaging} are mainly used to create 2D lattices. However, the CPW resonator unit cell size is limited by the wavelength and is sensitive to fabrication disorder due to their micro-meter scale CPW gap size. Moreover, the geometry and the nature of the distributed circuit elements in CPW resonator networks limit the implementation of a vast family of lattice structures, such as honeycomb or square lattices~\cite{kollar2019hyperbolic}. 

In this paper, we overcome this challenge by introducing novel superconducting lumped-element circuit architecture based on compact parallel-plate capacitors to demonstrate large-scale honeycomb lattices with flexible coupling connectivities. Inspired by the rotational symmetry-breaking mechanism on the honeycomb lattice~\cite{Yekta2023} for Dirac electrons, we introduce axially preferred second-neighbor coupling. When this coupling is the same for both sub-lattices the lattice realizes a tilting $\bs\zeta$ of the microwave cones along the direction of the added coupling~\cite{mota2021circuit} . To study the effect of the second-nearest neighbor coupling on the tilting angles of the dispersion relations, we characterize differently designed lattices of 731 sites at 15~mK temperature -- far below the superconducting critical temperature, where we realize narrow linewidth low-loss microwave modes-- and measure the DOS of the microwave collective modes to extract the tilting angles. In this work, we achieve a very strong tilting that matches the theoretical designed values of $59\%$  and $43\%$. To appreciate the significance of the above values, they should be compared with the tilting of the light cone by the gravity of the sun that was inferred from the deflection of the light path~\cite{Roy2019} and remains 5 orders of magnitude smaller than our tilt values. Our demonstration of controllable tilting of the microwave cone can be a basis for imprinting curvature into the fabric of the synthetic spacetime by allowing the tilt parameter $\bs\zeta$ to vary in space. This finding suggests superconducting circuits as promising platforms for emulation of aspects of  "general relativity in the lab" where one can study various properties of curved spacetime by variety of solid-state measurements. Enriching the solid-state platforms with the luxury of a curved background spacetime heralds possible new applications that arise from the structure of the synthetic spacetime geometry.

\section{Theory}

First we discuss how to synthetically incorporate tilting to the Dirac-like band crossing within our tight-binding model. The basic picture consists of adding anisotropic second neighbor hopping on a honeycomb lattice which can be conveniently realized by adding a second neighbor coupling only in one preferred direction denoted by red lines in the $y$-direction of Fig.\ref{fig:1}b.  The first neighbor coupling $J$ generates the parent (upright) microwave cone, while the selective second neighbor coupling $J'$ tilts it by $\zeta = 2J'/J$. This can be employed for controlled fabrication of a tilted microwave cone. The purpose of this paper is to experimentally demonstrate this concept.  The core idea is that a circuit based on a honeycomb lattice gives rise to a dispersion relation $\omega(\bs k)$ that around the band crossing points will be an upright microwave cone~\cite{mota2021circuit} in exactly the same way that an electron in $p_z$ orbitals hopping on the honeycomb lattice of graphene gives rise to a Dirac cone~\cite{Bostwick2006} (See Fig.~\ref{fig:1}a).  Consider the following Hamiltonian on the honeycomb lattice
\begin{equation}
    H = \sum_{i} \omega_r a^\dagger_i a_i + \sum_{i,\delta} J a^\dagger_{i+\delta}a_i + \sum_{i,\tau} J' a^\dagger_{i+\tau}a_i,
    \label{GenericHamiltonian.eqn}
\end{equation}
in which $\omega_r$ denotes the resonance frequency at $i$'th site. $\delta$ can be any of the three first neighbors connected by hopping $J$, whereas $\tau$ is the particular second neighbor denoted by red hopping $J'$ in Fig.~\ref{fig:1}b and is responsible for tilting the cone. To diagonalize the above Hamiltonian, one needs to Fourier transform to $k$-space. Since at every unit cell, there are two degrees of freedom associated with two-sublattices A (B) in Fig.~\ref{fig:1}a,b denoted by filled (empty) circles, after Fourier transformation one is left with a $2\times 2$ Bloch Hamiltonian that features band crossing at two points in the Brillouin zone~\cite{mota2021circuit}. This matrix can be expressed in terms of Pauli matrices $\sigma_x$ and $\sigma_y$ (see SI). Breaking the sublattice symmetry would introduce a perturbation proportional to $\sigma_z$ that would gap out the band crossing. Since here we do not wish to gap out the band crossing, we introduce the \emph{same coupling}  $J'$ for both AA and BB second neighbors. 

Since the second neighbor coupling is sub-lattice-diagonal, the off-diagonal Pauli matrices $\sigma_x$ and $\sigma_y$ are ruled out. On the other hand since both sub-lattices are equally involved, the Pauli matrix $\sigma_z$ is also ruled out~\cite{bernevig2013topo}. Hence, the only remaining choice to describe the effect of second neighbor coupling will be a term proportional to unit matrix $\sigma_0$.  This can be understood in terms of breaking the three-fold $C_{3v}$ symmetry of the honeycomb lattice down to two-fold $C_{2v}$ symmetry that arises from uniaxially preferred second neighbor coupling. In fact the irreducible representations of the $C_{2v}$ symmetry allow for a term in the Hamiltonian that is proportional to the unit matrix $\sigma_0$ and proportional to the scalar product of momentum $\bs k$ with some \emph{pseudo vector} $\bs\zeta$, i.e. a $\bs k\cdot\bs \zeta\sigma_0$ term which is precisely the tilting term~\cite{Goerbig2008}.  The idea of breaking $C_{3v}$ symmetry down to $C_{2v}$ symmetry to generate tilt in the cone is a generic mathematical fact and is equally valid for any lattice such as $8Pmmn$-borophene material~\cite{Yekta2023}.

The honeycomb lattices defined by couplings $J$ with symmetric first neighbor coupling lead to the crossing of the bands at two microwave cones indicated in Fig~.\ref{fig:1}c~\cite{bernevig2013topo}. In the circuit realization, each node is connected to all three nearest neighbors by inductance $L$ and selectively to only two second neighbors out of six by a $L'$ inductance. At every node, a capacitance $C$ is assumed to be connected to a common ground. Choosing the flux $\phi_j$ as the generalized coordinate, the Lagrangian will be given by~\cite{Girvin2014}
\begin{multline}
    {\cal L}=\frac{C}{2}\sum_j\dot\phi_j^2 \\
    -\frac{1}{2}\sum_j\left[\sum_{\delta}  \frac{(\phi_j-\phi_{j+\delta})^2}{2L}+\sum_{\tau}\frac{(\phi_j-\phi_{j+\tau})^2}{2L'} \right],
    \label{eqn.lagrangian}    
\end{multline}
where $\delta$ and $\tau$ denote the first and second neighbors. From the above Lagrangian,  the Hamiltonian in terms of generalized momenta $q_j=C\dot\phi_j$ (charge at node $j$) immediately follows
\begin{equation}
    H=\sum_j\left[\frac{q_j^2}{2C}+\frac{\phi_j^2}{2L_{\rm r}}\right]-\sum_j \left[\sum_\delta \frac{\phi_j\phi_{j+\delta}}{2L}+\sum_\tau\frac{\phi_j\phi_{j+\tau}}{2L'}  \right],
    \label{hamilfromlagrang.eqn}
\end{equation}
where $L_{\rm r}^{-1}=3L^{-1}+2L'^{-1}$.  Upon defining  $\omega_{\rm r}=1/\sqrt{L_{\rm r}C}$ and  $J = -\frac{1}{L}\sqrt{\frac{L_{\rm r}}{C}}$  and $J' = -\frac{1}{L'}\sqrt{\frac{L_{\rm r}}{C}}$,
the circuit model~\eqref{eqn.lagrangian} becomes a realization of the Hamiltonian~\eqref{GenericHamiltonian.eqn} that can be synthesized in the lab. Therefore, the tilting parameter $\zeta=2J'/J$ in our realization becomes~\cite{mota2021circuit}
\begin{equation}
    \zeta=\frac{2L}{L'}. \label{eqn.tilt}
\end{equation}
This profound formula shows that the dimensionless tilting of the microwave cones is simply tunable by adjusting the ratio of the first and second neighbor inductances in a circuit. Note that in the above derivation, we have neglected the boundary effects. The missing connections at the edge slightly modify $\omega_{\rm r}$. This must be considered in the numerical computations and interpretation of data for finite lattices. 
Euler-Lagrange equations of motion for the above Lagrangian are equivalent to Kirchhoff's law (c.f. SI). Due to the second time derivative which is a characteristic of the bosonic Harmonic oscillators, the eigenvalues of the equation of motion will be $\omega^2/\omega_0^2$ with $\omega_0^2=1/(LC)$, where $\omega$ are resonance frequencies obtained from the Hamiltonian~\eqref{hamilfromlagrang.eqn}.  It has been shown~\cite{mota2021circuit} that the above Hamiltonian features a tilted microwave cone where the tilt parameter is given by $\zeta$ in Eq.~\eqref{eqn.tilt} (please see SI for more details). 



\begin{figure*}[t]
	\centering
	\includegraphics[width = \textwidth]{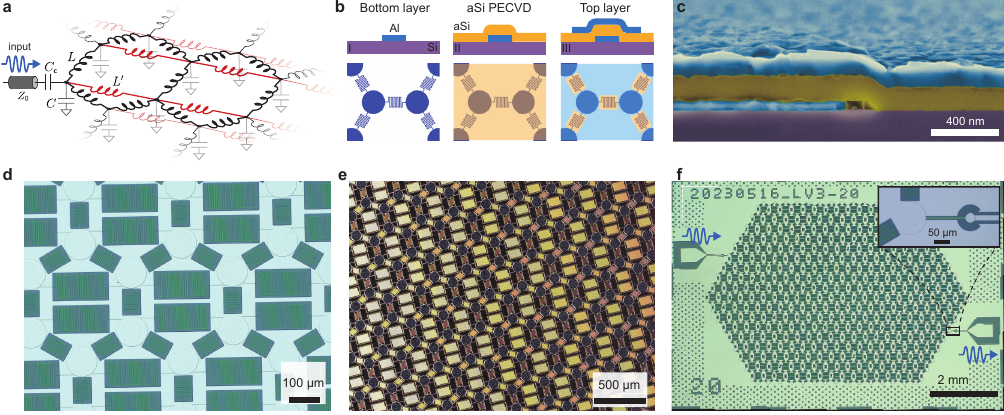}
	\caption{\textbf{Large-scale superconducting circuit lattices.} \textbf{a}, The circuit diagram of a honeycomb lattice with uniaxial second nearest neighbor coupling. Nearest neighbor nodes are connected with lumped element inductance $L$, while the second nearest neighbors are coupled with inductance $L'$. Each node is grounded through a constant capacitance $C$. The finite circuit lattice is weakly coupled to the external measurement circuit chain through a coupling capacitance $C_\mathrm{c}$ followed by an impedance-matched waveguide. \textbf{b}, Nanofabrication process flow for realization of our superconducting circuit lattice. The inductors and bottom plates of capacitors are lithographically defined with $75$~nm of Aluminum. A $200$~nm dielectric layer of aSi is deposited and covered by a $150$~nm Al layer defining the top capacitor plates and the ground plane. \textbf{c}, False-colored SEM image showing a capacitor cross-section. \textbf{d}-\textbf{e}, Optical micrographs showing the honeycomb superconducting circuit lattice. \textbf{f}, Optical image showing a full chip with $731$ sites. The chip is coupled on two sides to co-planar waveguides through the coupling capacitors shown in the inset.}
    \label{fig:2}
\end{figure*}

The general-relativistic significance of a tilt $\bs\zeta$ in the dispersion relation is that for infinite system the dispersion relation can be represented as $g^{\mu\nu}k_\mu k_\nu=0$~\cite{Volovik2021,Jafari2023} where $k_\mu=(-\omega/v,k_x,k_y)$. The emergent \emph{spacetime metric} $g_{\mu\nu}$  defined by $ds^2=-v^2dt^2+(d\mathbf{x}-v\bs\zeta dt)^2$ is precisely the matrix inverse of the above $g^{\mu\nu}$ that arises from the tilted Dirac-like dispersion relation~\cite{Farajollahpour2019,JalaliMola2019,Jafari2019,Farajollahpour2020}. For details please refer to SI. Note that the above metric can be obtained from the metric $ds^2=-v^2dt^2+d\mathbf{x}^2$ of upright Dirac cone, by transformation $t\to t$, $\bs x\to \bs x-\bs\zeta vt$. This transformation allows us to attribute the difference between the group velocities  $v^+_g$ and  $v^-_g$  in Fig.~\ref{fig:1}.b to a moving frame~\cite{Jafari2023}. In another words, converting an upright cone to a tilted cone is a solid-state realization of a moving frame.


\section{Fabrication of superconducting circuit lattice }

In order to experimentally implement the circuit model in Eq.~\eqref{hamilfromlagrang.eqn}, we introduce a novel superconducting circuit lattice architecture based on the lumped element parallel-plate capacitors and planar meander inductors. Figure~\ref{fig:2}a shows the circuit diagram of the lattice, where each node on the honeycomb structure is connected to a common ground by a capacitor ($C$), and the first and second nearest neighbor nodes are connected via inductors, $L$ and $L'$ respectively. A finite lattice can be coupled to a measurement circuit chain through a coupling capacitor $C_\mathrm{c}$ connected to a 50-$\mathrm{\Omega}$ impedance-matched waveguide. As shown in Fig.\ref{fig:2}c, to realize such a lattice, first, we lithographically define meander inductors and bottom plates of capacitors with 75~nm sputtered aluminum on a high-resistivity silicon substrate. Next, we deposit a 200~nm amorphous silicon at 200°C using PECVD to ensure perfect step coverage. To reduce the capacitors' footprint, amorphous silicon is chosen as the dielectric material due to its relatively low microwave losses~\cite{o2008microwave} and high dielectric coefficient ($\epsilon_\mathrm{r} \simeq 11.7$). Finally, we deposit the top Al layer defining the ground plane and covering the entire circuit, except for meander inductors. This minimizes undesirable parasitic couplings between adjacent inductors. Figures~\ref{fig:2}~d, e, and f show microscope images of the fabricated device with 731 nodes. Cross-section SEM images show perfect step coverage of the dielectric layer in the capacitors (Fig.\ref{fig:2}c).
The coupling ports of the lattice are shown in the inset of Fig.\ref{fig:2}f. The value of coupling capacitance (Fig.\ref{fig:2}f inset) is chosen to maximize transmission between two ports of the lattice (See SI). To reduce the impact of edge modes, we simulated the best nodes to couple input and output ports on the edges before the fabrication (see SI).
Compared to CPW lattices~\cite{underwood2012low}, the present lumped element circuit architecture results in a superconducting circuit with 1000-fold smaller form factors of only 0.02~mm$^2$ footprint. This enables the realization of large-scale and low-disorder circuit lattices.
In this work, we designed three devices with identical $L=0.37$ nH and $C=8.1$ pF. In one device there is no second neighbor inductor whereas the other two have $L' = 1.9$~nH, and $1.4$~nH. According to Eq.\eqref{eqn.tilt}, three devices are designed to have tilt parameters of $\zeta = 0, 0.40, 0.52$.

\begin{figure*}[t]
	\centering
	\includegraphics[width = \textwidth]{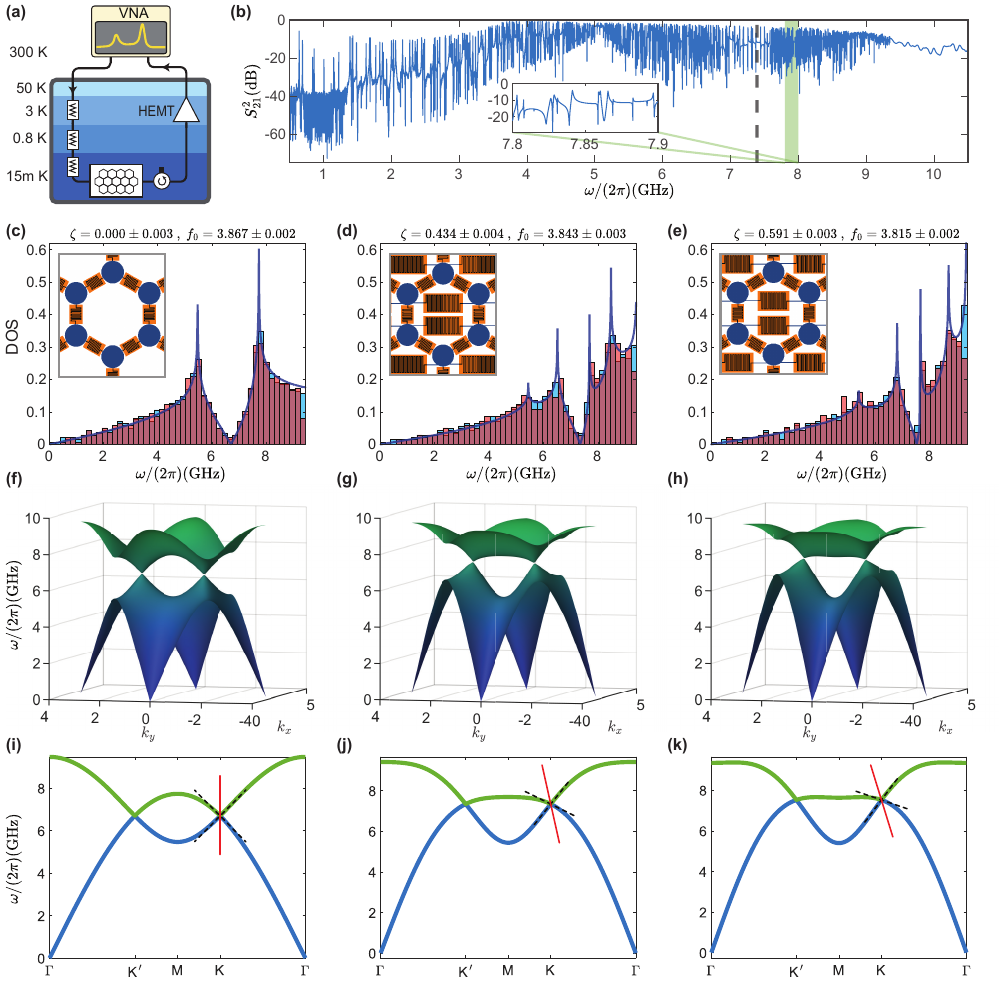}
	\caption{\textbf{Observation of the band structure and tilted Dirac-like microwave cones.} \textbf{a}, Experimental setup for the measurement of scattering parameter $S_{21}$. The superconducting device is operated in a dilution refrigerator with a base temperature of $15$~mK.  \textbf{b}, The plot of measured transmission coefficient $|S_{21}|^2$ of the lattice with $L'= 1.9$~nH as a function of frequency. The inset magnifies the data of the indicated green-shaded frequency range.  \textbf{c}-\textbf{e},  Experimentally extracted (blue bar) and theoretically computed (orange bar) DOS of devices with different $\zeta$ parameters. The green curve is the DOS of infinite lattice for the indicated numerically fitted parameters demonstrating the substantial lattice size of the fabricated devices.  The insets show the corresponding circuit layout.  \textbf{f}-\textbf{h}, Reconstructed microwave band structure of each device. \textbf{i}-\textbf{k}, 2D cross-section of the band structure along KK' line emphasizing the tilting of the Dirac-like microwave cone. The tangents indicate the cone's lateral side, and the red solid axis indicates the tilting.}\label{fig:3}
\end{figure*}

\section{Experimental Results}
In order to extract the spectrum of the device, we measure the microwave transmission scattering parameter $S_{21}$ through its two coupling ports. Figure \ref{fig:3}a schematically shows the measurement chain consisting of a microwave vector network analyzer, cryogenic attenuators, a cryogenic amplifier, and a microwave isolator. The device is operated at the $15$ mK stage of a dilution refrigerator.

The measured transmission coefficient $|S_{21}|^2$ for  the device with $L' = 1.9$~nH is displayed in Fig.~\ref{fig:3}b.  The inset shows that individual resonances are well separated. To automate the detection of resonance peaks, we have developed an algorithm based on phase gradients (PhG) and amplitude of the recorded trace $S_{21}$ (See SI). In this way we are able to detect $98\%$ of the modes prior to manual verification. 

The DOS of the devices is constructed from identified resonance frequencies shown in Figs.~\ref{fig:3}~c-e with blue bar plots. The actual values of the parameters $\omega_0$ and $\zeta$ for each device are obtained by fitting the numerically computed DOS for our theoretical model on a finite lattice to the measured DOS. These are denoted by orange bar plots in Fig.~\ref{fig:3}c-e in good agreement with the measured DOS (blue bars). The curved plot is the DOS of the infinite lattice corresponding to the fitted values of $\omega_0$ and $\zeta$. The good matching between the measured DOS and the infinite lattice DOS is a result of large number of sites that are achievable in our platform. 

 The DOS for both finite and infinite lattice feature  two sharp features known as van Hove singularities on two sides of the Dirac-like  crossing point which arise from saddle points in the dispersion relation. This aspect is  similar to the van Hove singularities in graphene with $\zeta=0$ . However, the DOS for the circuit lattice is not symmetric as in graphene. The asymmetry is due to the fact that the eigenvalues of the admittance matrix of circuit lattice are $(\omega/\omega_0)^2$~\cite{mota2021circuit}.  Upon increasing $\zeta$, as can be seen by comparing Figs.~\ref{fig:3} c-e, this asymmetry is enhanced. Furthermore, additional van Hove singularities in the DOS appear for non-zero $\zeta$ . In Figs.\ref{fig:3}f-h the band structure of the infinite lattice corresponding to fitted values of $\omega_0$ and $\zeta$ is plotted as a function of two-dimensional $\vec{k}$ in the 1st Brillouin zone. Finally, in Figs.~\ref{fig:3}i-k the cross-section of the 3D band structure along the KK' direction is shown, exhibiting stronger tilt for larger $\zeta$ guided by solid red lines at the crossing point. Note that the effect of the parameter $\zeta$ is not limited to tilting at frequencies around the crossing point. As can be seen in both DOS and band structure plots, the effect of second neighbor coupling is not limited to tilting of the features around the Dirac-like crossing point. Frequencies away from the crossing point are also substantially affected by the $\zeta$. Recently it has been shown that the bending of the energy scales above the Dirac node in fermionic tilted Dirac cones are essential in producing smart black holes capable of correctly reproducing temperature \textit{and} entropy of the general-relativistic black holes~\cite{Afshordi2024}. 

\section{Conclusion and outlook}
In this work, we introduced a novel experimental platform to demonstrate deterministic tilting of a Dirac-like microwave band crossing by desired amount. Harnessing the tilting of a microwave band structure is tantamount to engineering the fabric of the ensuing emergent spacetime geometry. Spatial/temporal manipulation of circuit parameters of the lattice results in $\boldsymbol{\zeta}(\boldsymbol{x},t)$, thereby inducing the Painlev\'e-Gulstrand (PG) family of spacetime metric $ds^2=-v^2dt^2+\left(d\boldsymbol{x}-v\boldsymbol{\zeta}(\boldsymbol{x},t)dt\right)^2$~\cite{Martel2001}. Since $v\boldsymbol{\zeta}$ represents a moving frame velocity, a variable $\boldsymbol{\zeta}$ will correspond to acceleration which according to general relativity principles correspond to a curved spacetime geometry. Examples of gravity-like effects arising from $\boldsymbol{x}$- and/or $t$-dependent $\boldsymbol{\zeta}$ that can be studies in superconducting circuits include: (1)  \emph{Time dilation}: One manifestation of this effect would be enhancement of the DOS slope around the band crossing point -- similar to that in Figs.~\ref{fig:3}c-e. This immediately follows from the fact that tilting a cone changes the constant energy surface from circle to ellipse whose area is enhanced by $1/\sqrt{1-\zeta^2}$ ~\cite{mota2021circuit}.  The same enhancement  is expected to appear in the measurements of "time" intervals. Setting $d\boldsymbol{x}=0$ in the invariant distance $ds^2\equiv -d\tau^2$  leads to $dt=d\tau/\sqrt{1-\zeta^2}$. This means that time intervals depend on the local value of $\zeta$. Note that the "time" here does not refer to the time measured by the clocks in the lab, rather, it is defined by the frequency of the probe LC resonator at a particular point of the lattice that plays the role of coordinate clock in the relativity. A smoking gun evidence of underlying spacetime metric would be to use identical probe LC resonators to measure the dependence of their frequencies on $\zeta$ in the different devices or different areas of the same device with spatially variable $\boldsymbol{\zeta}$.  Due to moving frame interpretation of the tilt parameter, this resembles the different rates of clocks in moving spaceships with different speeds. (2) \emph{Gravitomagnetic effects}: When the $\bs\zeta$ starts to depend on space and/or time, we will have a solid state realization of the corresponding PG spacetime. Creating a vortex profile for $\bs\zeta$ with non-zero $\bs\nabla\times\bs\zeta=\bs\omega$ will generate a geometry that looks like a rotating gravitational source~\cite{Rostamzadeh2023,Hosseinzadeh2023,Farajollahpour2020}.  One way to achieve the above vortex profile in the background PG spacetime is to implement a gradient in $x$ direction in the circuit elements of our devices featuring a tilt in the $y$ direction. The above tilt profile is a "gravito"-magnetic field~\cite{Hosseinzadeh2023,Rostamzadeh2023}, thereby cyclotron-like orbits in the classical limit and Landau-quantization behavior with characteristic $\sqrt{n}$ energy levels for the microwaves are expected. This can serve as a smoking gun evidence of gravito-magnetic effects in circuits. (3) \emph{Generation of "gravitational" waves}: Josephson-based flux tunable inductors can be used to imprint a spatial profile on the circuit parameters by magnetic fields that amounts to non-uniformity in the PG spacetime parameters.  Even for a spatially uniform flux, driving it with frequency $\omega_{\rm gw}$ drives the PG spacetime at the same frequency, and hence can be regarded as a way of generation of "gravitational waves" at least when $\omega_{\rm gw}$ is smaller than the natural frequency $\omega_{\rm r}$ of the circuit.  (4) \emph{Quantum behavior and the curved spacetime:} Our platform becomes a viable route to integrate superconducting qubits with large scale circuit lattices to study  the role of "curved spacetime" in mediation of entanglement between the qubits~\cite{Deli2020,AspelmeyerAvoidClassical,Christodoulou2023} . 


\subsection*{Acknowledgment}
This work was supported by the European Research Council (ERC) grant No. 835329 (ExCOM-cCEO). This work was also supported by the Swiss National Science Foundation (SNSF) under grant No. NCCR-QSIT: 51NF40\_185902 and No. 204927. S.A.J. was supported by Alexander von Humboldt foundation. The contribution of A.M. in this work has been submitted as part of his PhD thesis. All devices were fabricated in the Center of Micro-Nano Technology (CMi) at EPFL. We thank Mahdi Chegnizadeh for his support in conducting the cryogenic experiment.

\textbf{Authors contribution}

S.A.J. conceived the project and led the theoretical line of thought. A.Y. and A.M. designed and simulated the superconducting circuit.  A.Y. fabricated the devices and performed the measurements with support from T.J.K. A.M. performed the theory computations and analyzed the data. S.K. critically read the manuscript and contributed to the writing of the paper. The paper was discussed and jointly written by A.M., A.Y., S.K., and S.A.J.

\textbf{Data and code availability}

The data and codes used to produce the plots within this paper will be available on Zenodo.
All other data used in this study are available from the corresponding author upon reasonable request.



\putbib[Refs]
\pagebreak
\end{bibunit}

\newpage

\begin{bibunit}[plain]

\begin{widetext}
\begin{center}
{\bf Supplementary Information for: Realization of tilted Dirac-like microwave cone in superconducting circuit lattices}
\end{center}



\setcounter{equation}{0}
\renewcommand{\theequation}{S\arabic{equation}}
\renewcommand{\thefigure}{S\arabic{figure}}
\renewcommand{\theHfigure}{S\arabic{figure}}
\setcounter{figure}{0}
\setcounter{table}{0}

\setcounter{subsection}{0}
\setcounter{section}{0}

\tableofcontents


In this supplement, in addition to providing pedagogical introduction to the concepts from special/general relativity that can be followed with basic solid-state physics background, we offer details of the computations to enable a self-contained understanding of the main text. 

\section{From dispersion relation to spacetime metric}
In this paper we draw implications for the structure of underlying "spacetime" by looking into the dispersion relation that gives energy of excitations as a function of their wave-vector $\bs k$. It is important to make a connection between the dispersion relation of excitations and the geometry of the spacetime. In this section we motivate how the tilting of dispersion relation can be related to a non-trivial spacetime metric.  The upper limit of speeds for fundamental particles that are excitations in the vacuum  is given by the speed of light $c$, whereas in the solid-state it is given by another velocity scale that associates an energy scale with hoppings to neighboring lattice sites. For simplicity we set the velocity scale to unit in the following discussion to emphasize that the mathematical structure is the same for fundamental particles in the vacuum and excitations on a lattice. 

In three space dimensions the space and time coordinates can be combined into a single spacetime vector called a four-vector $x^\mu=(t,\bs x)$ where $\bs x=(x,y,z)$. Likewise in two space dimensions where  $\bs x=(x,y)$, the spacetime vector $x^\mu$ with $\mu=0,1,2$ will denote a "three vector" in the spacetime.  Corresponding to the above spacetime coordinates $x^\mu$ one can define components of a gradient vector $\partial_\mu\equiv \partial/\partial x^\mu$. We know that quantum mechanically the time derivative $\partial_0=\partial/\partial t$ is related to energy and $\nabla=\partial/\partial \bs x$ is related to momentum. Therefore (setting $\hbar=1$ so that momentum $\bs p=\hbar\bs k$ will be the same as wave vector $\bs k$) it is natural to combine energy and momentum to form another spacetime vector $k_\mu=(-\varepsilon,\bs k)$. To see where does this notation come from, let us start with the energy-momentum dispersion relation for upright Dirac cone, i.e. $\varepsilon=\pm\sqrt{|\bs k|^2+m^2}$ where the mass $m$ could be zero or non-zero. One can write it equivalently as $-\varepsilon^2+\bs k^2=-m^2$. This relation can be cast into matrix form 
\begin{equation}
    (-\varepsilon,k_1,k_2)\begin{pmatrix}
    -1 & 0 & 0\\
    0  & 1 & 0\\
    0  & 0 & 1
    \end{pmatrix} 
    \begin{pmatrix}
    -\varepsilon \\ k_1 \\ k_2
    \end{pmatrix}
    =-m^2.
\end{equation}
Using the shorthand notation $k_\mu=(-\varepsilon,\bs k)$ and denoting the $3\times 3$ matrix by $\eta^{\mu\nu}={\rm diag}(-1,1,1)$, the above equation can be written in the covariant (and compact) form
\begin{equation}
    \sum_{\mu\nu}k_\mu \eta^{\mu\nu} k_\nu=-m^2 \to k_\mu \eta^{\mu\nu} k_\nu=-m^2
    \label{Squadratic.eqn},
\end{equation}
where we have used Einstein's summation convention that implies a summation over repeated indices that appear as both subscript and superscript. This allows to save in writing $\sum$ symbols. The set of transformations 
\begin{equation}
    \begin{pmatrix}
    -\varepsilon' \\ k_1' \\ k_2'
    \end{pmatrix} = \begin{pmatrix}
    \Lambda_0^0 & \Lambda_0^1 & \Lambda_0^2\\
    \Lambda_1^0 & \Lambda_1^1 & \Lambda_1^2\\
    \Lambda_2^0 & \Lambda_2^1 & \Lambda_2^2
    \end{pmatrix}
    \begin{pmatrix}
    -\varepsilon \\ k_1 \\ k_2
    \end{pmatrix}
\end{equation}
or in compact form $k'_\mu=\Lambda_\mu^\nu k_\nu$ that leave the left side of Eq.~\eqref{Squadratic.eqn} invariant, i.e. lead to the equation $k'_\mu \eta^{\mu\nu} k'_\nu=-m^2$ are called Lorentz transformation~\cite{SIRyder2009}. Therefore the Dirac type dispersion relation is intimately related to the metric $\eta^{\mu\nu}={\rm diag}(-1,1,1)$ of the 1+2 dimensional Minkowski spacetime. \emph{This is how the dispersion relation of the excitations informs about the geometry of the underlying spacetime structure. } Therefore it is not surprising that the tilting of the Dirac cone can be accommodated by adding extra entries to the spacetime metric that defines the length of spacetime energy-momentum vectors. 

To work out the spacetime metric that corresponds to a tilted Dirac cone, let us simply start from the dispersion relation of a tilted Dirac cone $\varepsilon=\pm\sqrt{\bs k^2+m^2}+\bs\zeta.\bs k$ where $\bs\zeta$ is the amount of tilting. It can be rearranged into the form $-(\varepsilon-\bs\zeta\bs k)^2+\bs k^2=-m^2$. Expanding the left hand side and using the spacetime vector notation $k_\mu=(-\varepsilon,\bs k)$ the dispersion relation can be cast into the form $g^{\mu\nu}k_\mu k_\nu=-m^2$ where 
\begin{equation}
    g^{\mu\nu}=\begin{pmatrix}
        -1 & -\zeta_x & -\zeta_y\\
        -\zeta_x & 1-\zeta_x^2 & -\zeta_x\zeta_y\\
        -\zeta_y & -\zeta_y\zeta_x & 1-\zeta_y^2
    \end{pmatrix},
\end{equation}
 defines a metric to measure the length of energy-momentum vectors using their covariant components $k_\mu$ as $g^{\mu\nu}k_\mu k_\nu$. Inverting the above matrix gives the metric components $g_{\mu\nu}$ that can be used to evaluate the length of spacetime vectors $x^\mu$ as $g_{\mu\nu}x^\mu x^\nu$ where 
\begin{equation}
    g_{\mu\nu}=\begin{pmatrix}
        -1+\zeta^2 & -\zeta_x & -\zeta_y\\
        -\zeta_x & 1 & 0\\
        -\zeta_y & 0 & 1
    \end{pmatrix}.
\end{equation}
Using the above metric,  length of an infinitesimal spacetime vector $dx^\mu$ is $ds^2=g_{\mu\nu}dx^\mu dx^\nu$. After using the above matrix and simplification it becomes $ds^2=-dt^2+(d\bs x-\bs\zeta dt)^2$ which is a compact representation of the spacetime that corresponds to a tilted Dirac cone. In fact setting $\bs\zeta=0$ gives the familiar length element $-dt^2+d\bs x^2$ of the Minkowski spacetime that corresponds to the dispersion relation of an upright Dirac cone.  This reveals a simple but profound fact: The metric of a tilted Dirac cone can be obtained from the metric of the upright Dirac cone by the "moving frame" transformation 
\begin{equation}
    t\to t,~~~\bs x\to \bs x-\bs\zeta t.
\end{equation}
This means that the spacetime of a tilted Dirac cone is like a moving frame with respect to an upright Dirac cone. The ability to make tilting parameter or frame velocity $\bs\zeta$ to depend on space and/or time will be a variable velocity and hence an acceleration. On the other hand, according to principles of general relativity, accelerated frame is equivalent to a spacetime geometry. This is how tilt parameters as moving frame velocity will be able to imprint a spacetime geometry on our circuit lattice. The above argument relies only on the conic shape of the dispersion relation. In the case of Fermions such cones are called Dirac cones. In the case of visible light it is called light cone. In the case of circuit resonators, microwave cone would be an appropriate name for the Dirac-like dispersion. We use both Dirac (-like) and microwave cones interchangeably. 

\section{Theory of tilted microwave cone in circuit resonator graphs}
The structure of the circuit lattice is shown in Fig.~\ref{fig:sup_fig_1}. Left and right parts indicate top and side view. Each node or site is connected to its first three neighbors via the base inductance $L$ and to the common ground by a capacitor $C$. If these were the only edges on the honeycomb lattice, one would obtain upright Dirac cone~\cite{SImota2021circuit} similar to the case of Graphene.  To externally impose a preferred axis in the plane of honeycomb lattice, we connect each node to its second neighbors in the horizontal ($y$) direction by an inductance  $L'$. The other four second neighbors links are not present. This procedure promotes the honeycomb lattice into honeycomb graph. This essential step is responsible for tilting of the resulting microwave cone. The unit cell which is shown in the top view by dashed lines contains two nodes (belonging to $A$ and $B$ sub-lattices, respectively). The structure is repeated in both $x$ and $y$ directions to ensure a two-dimensional nature. Theoretically it can be subject to open or periodic boundary condition, whereas in the lab open boundary conditions will be realized.
\begin{figure*}[t]
	\centering
	\includegraphics[width = 1\textwidth]{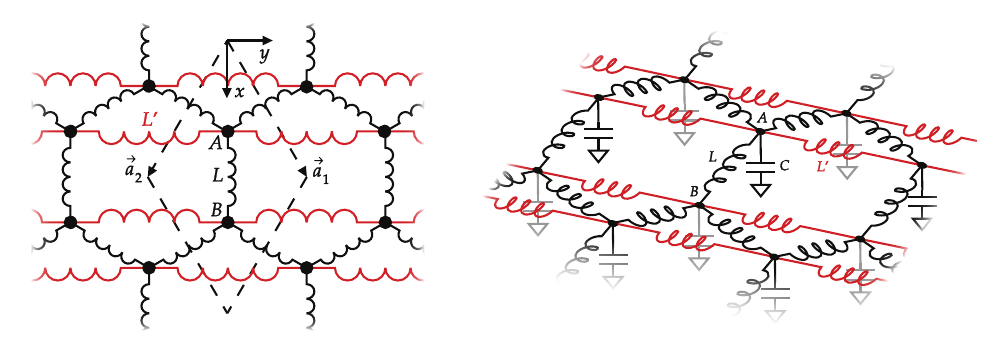}
    \caption{Top and side view of the honeycomb graph formed by circuit elements.}
    \label{fig:sup_fig_1}
\end{figure*}


In order to find the equations of the circuit we start from the Lagrangian we introduced in the main text~\ref{eqn.lagrangian}. We use Euler-Lagrange equations to find the differential equations for the voltages at each node:
\begin{equation*}
	\frac{d}{dt} \frac{\partial \mathcal{L}}{\partial \dot{\phi_j}} = \frac{\partial \mathcal{L}}{\partial \phi_j}
\end{equation*}
Substituting the Lagrangian in the above equation we get
\begin{equation*}
	C \dot{{\bf \phi}_i} = - \left[ \sum_{\delta} \frac{\phi_i - \phi_{i+\delta}}{L}  + \sum_{\tau} \frac{\phi_i - \phi_{i+\tau}}{L'} \right].
\end{equation*}
In this equation $\delta$ runs over all first neighbors'  and $\tau$ shows second neighbors'  which is limited only to the $\pm(\bs{a}_1-\bs{a}_2)$ second-neighbors of the $i$th site. Note that the left side is the current through the capacitor connected to node $i$ and the right side is the sum of currents through this node connected to inductors. Therefore these are the Kirchhoff current law (KCL) equations for each node. Using time harmonic solutions for each node's voltage we get
\begin{equation}
\sum_{\delta} \frac{V_{i}-V_{i+\delta}}{\ii \omega L} + \sum_{\tau} \frac{V_{i}-V_{i+\tau}}{\ii \omega L'} + \ii\omega C V_{i} = I_{i}.
\label{kcl_eq}    
\end{equation}   
 The sum over $\tau$ of the second neighbors encodes the uni-axial preference at the lattice level by excluding the other four second-neighbors $\pm\bs a_1$ and $\pm\bs a_2$. To account for the open boundary condition in the experiment, when this site is located on the (open) boundary of the lattice, corresponding first or second neighbors will be excluded from the sum. We consider this point when we numerically solve the problem for a finite lattice. Also note that, in this equation we assume there could be an external current source $I_i$ injecting at node $i$ that can be used for the excitation of the circuit by external current. For example if we want to send a signal to the lattice via a specific site or measure the output signal from another site we set these site's current to a desired nonzero value. To find the natural resonances of the lattice, no probe external currents are required and we set the values of all the external currents to zero. In this situation all the KCL equations for lattice sites can be written in a matrix equation,
\begin{equation*}
Y 
\begin{pmatrix}
    \vdots\\
	V_{i}\\
	\vdots\\
\end{pmatrix} = \textbf{0}.
\end{equation*}
In this equation $Y$ is the admittance matrix of the lattice which encompasses the admittance connectivity of the sites. Nonzero elements in its $i$'th row are the values related to the KCL equation for this site. Note that \textbf{0} is a zero vector whose length is equal to the number of sites. By multiplying $\ii\omega L$ to the equation we get a much simpler equation for each row.
\begin{equation*}
	\sum_{\delta} (V_{i}-V_{i+\delta}) + \frac{\zeta}{2} \sum_{\tau} (V_{i}-V_{i+\tau}) = \left(\frac{\omega}{\omega_0}\right)^2 V_{i},
\end{equation*}
where $\omega_0=1/\sqrt{LC}$ and $\zeta = 2L/L'$. These new equations form a matrix equation $D V = (\omega/\omega_0)^2 V$ where $D$ is a Hermitian dynamical matrix. This equation is an eigenvalue problem with eigenvalues $\lambda = (\omega/\omega_0)^2$. We denote the the $j$'th eigenvalue and its corresponding normalized eigenvector by $\lambda_j$ and $V_{j}$ which is column vector with $N_s$ (number of sites) components. We define $\Gamma$ as the diagonal matrix of eigenvalues which is real and $\tilde{V}$ as the matrix consisting of eigenvector $V_j$ which is a unitary matrix. Therefore we get the relation $D = \tilde{V} \Gamma \tilde{V}^{\dagger}$. Note that the relation between $D$ and $Y$ is
\begin{equation}
    \ii \omega L Y = D - \lambda \mathbb{I}
    \label{YD_rel}
\end{equation}
where $\mathbb{I}$ is the identity matrix.

\subsection{Fourier analysis for infinite lattice}
For a lattice with periodic boundary condition the discrete transnational symmetry between unit cells denoted in Fig.~\ref{fig:sup_fig_1} allows us to use 2D discrete Fourier transform to reduce the eigenvalue problem to a two dimensional matrix eigenvalue problem in the momentum space that can immediately give the band structure. Each unit cell has two sites belonging to sub-lattices A and B. Here we use another subscript to specify the sublattice. Therefore $V_{i s}$ is the voltage on unit cell $i$ and sublattice $s$ which can be either A or B. By employing discrete translation symmetry via Fourier transform, we write $V_{i s} = \sum_{\bs{k}} V_{\bs{k}s} e^{\ii \bs{k}.\bs{r}_i}$ in which $\bs{k} = (k_x,k_y)$ is the wave vector in the $\bs{k}$-space and $\bs{r}_i$ shows the two dimensional position of each unit cell. By scaling $\bs{k}$ the position label of the nodes can be correspondingly changed, so we can set $\bs{a}_{1(2)} = a_0(\sqrt{3},\pm 1)/2$ in which $a_0$ is distance between second neighbors. Also, the reciprocal lattice vectors are $\bs{b}_{1(2)} = 2\pi(1/\sqrt{3},\pm 1)/a_0$. Note that since the frequency (energy) scales are determined by the capacitance $C$ and the inductance $L,L'$, the parameter $a_0$ serves only for labeling and has no effect on the dispersion relation, so it can be assume to be the unit. When we are on node $A(B)$ the first neighbor distances are given by shifting  by the bond length of $a_0/\sqrt 3$ along either of the three bonds of honeycomb lattice. Defining $\bs\delta=(1,0)a_0/\sqrt{3}$ the three first neighbors are given by 
\begin{equation*}
    \bs\delta_1 = \bs\delta ~,~ \bs\delta_2 = \mp\bs{a}_1+\bs\delta ~,~ \bs\delta_3 = \mp\bs{a}_2+\bs\delta.
\end{equation*}
In this basis, our chosen second neighbor vectors are given by 
\begin{equation*}
    \bs\tau = \pm (\bs{a}_1 - \bs{a}_2).
\end{equation*}
Applying Fourier transform to the matrix equation we get 
\begin{equation}
    \begin{pmatrix}
        \epsilon(\bs{k}) & \Delta(\bs{k}) \\
        \Delta^*(\bs{k}) & \epsilon(\bs{k})
    \end{pmatrix} 
    \begin{pmatrix}
        V_{\bs{k}A} \\
        V_{\bs{k}B}
    \end{pmatrix} = 
    \left(\frac{\omega}{\omega_0}\right)^2
    \begin{pmatrix}
        V_{\bs{k}A} \\
        V_{\bs{k}B}
    \end{pmatrix},
    \label{dynamical_eig_k}
\end{equation}
where 
\begin{align*}
\epsilon(\bs{k}) & = 3 + \zeta (1-\cos{\bs{k}.(\bs{a}_1 - \bs{a}_2})), \\
\Delta(\bs{k}) & = -1 - e^{-\ii\bs{k}.\bs{a}_1} - e^{-\ii\bs{k}.\bs{a}_2}.  
\end{align*}
Solving this two by two matrix eigenvalue problem we have two frequency bands with values
\begin{equation*}
    \omega_{\pm} = \omega_0 \sqrt{\epsilon \pm |\Delta|}.
\end{equation*}
When $\Delta = 0$ the the two bands touch each other. To prove that this defines a Dirac-like band crossing, we need to show that around this points the dispersion relation is linear like Dirac materials. In $\bs{k}$-space the location of these points are obtained from these conditions
\begin{align*}
    1 + \cos{\bs{k}_D.\bs{a}_1} + \cos{\bs{k}_D.\bs{a}_2} & =  0, \\
    \sin{\bs{k}_D.\bs{a}_1} + \sin\bs{k}_D.\bs{a}_2 & = 0.
\end{align*}
The only possible solution is when $\bs{k}_D.(\bs{a}_1 + \bs{a}_2) = 2m\pi$  and $\bs{k}_D.\bs{a}_1 = \pm 2\pi/3 + 2n\pi$ ($m,n \in \mathbb{N}$) which result in
\begin{equation*}
    k_{Dx} = \frac{2m\pi}{\sqrt{3}a_0} ~,~ k_{Dy} = \pm\frac{4\pi}{3a_0} + (2n-m)\frac{2\pi}{a_0}.
\end{equation*}
There are two distinct Dirac-like points (two valleys) in 1st Brillouin zone given by $\bs{K}(\bs{K}') = 2\pi/(3a_0)(\sqrt{3},\nu)$ where $\nu = \pm 1$. The resonance frequency at Dirac points is
\begin{equation}
    \omega_D = \omega_0 \sqrt{3(1+\zeta/2)}.
    \label{eq: Dirac_frquency}
\end{equation}
In the next step we expand the band dispersion around Dirac points by considering the resonance frequency $\omega = \omega_D + \delta\omega$ and wave vector $\bs{k} = \bs{k}_D + \delta\bs{k}$. Now the matrix elements of Eq. \ref{dynamical_eig_k} take the form
\begin{align*}
\epsilon(\bs{k}) & = 3(1+\zeta/2) + \sqrt{3} \nu \frac{\zeta}{2} \delta\bs{k}.(\bs{a}_1 - \bs{a}_2) = 3(1+\zeta/2) + \sqrt{3} \nu \frac{\zeta}{2} a_0 \delta k_y \\
\Delta(\bs{k}) & = -\frac{\ii}{2} \delta\bs{k}.(\bs{a}_1+\bs{a}_2) + \tau \frac{\sqrt{3}}{2} \delta\bs{k}.(\bs{a}_1-\bs{a}_2) = \frac{\sqrt{3}}{2} a_0 (-\ii\delta k_x + \tau \delta k_y).      
\end{align*}
Therefore the resonance frequency expansion around Dirac points is
\begin{equation*}
    \frac{\delta \omega_\pm}{\omega_0} = \frac{1}{4\sqrt{1+\zeta/2}}\left(\nu \zeta \delta k_y \pm \sqrt{\delta k_x^2 + \delta k_y^2}\right)
\end{equation*}
which is a linear dispersion relation with a cone shape in $\omega-\bs{k}$ space. The first term shows tilting of the Dirac cone along the $\nu y=\pm y$ directions for valleys $\nu=\pm$ corresponding two Dirac points $\bs{K}$ and $\bs{K}'$.  Therefore the uni-axial preference incorporated by the inductance $L'$  along $\pm y$ axis has lead to the tilting of one of the Dirac cones, one along $+y$ and the other one along the $-y$ direction. 

\subsection{Coupling finite lattice to input and output ports}
In order to measure resonance frequencies of the finite lattice we need to locally excite the circuit lattice. We do this at a site on one of the borders of the lattice in a range of frequencies and measure the output signal at another site in the other side of the lattice. The ratio of output voltage ($V_{\rm out}^+$) to input voltage ($V_{\rm in}^+$) is called the transmission scattering matrix $S_{21}$. There is also a reflection from input ($V_{\rm in}^-$) and its ratio to input voltage is called reflection scattering matrix element $S_{11}$. By sweeping a range of input frequencies, at certain resonances (eigen-frequencies) the lattice lets the signal to pass through, leading to sharp peaks in $S_{21}$ at such frequencies. In this subsection we compute the dependence of scattering matrix to the admittance of the lattice and the impedance of input or output coupling elements. In the next section we use this relation to determine the best range for the values of coupling parameters in order to attain sharp peaks on $S_{21}$ with small line-width. In another words we require the resonances to have high quality factors. 

Consider a finite honeycomb lattice of inductors and capacitor elements with admittance $Y$ which has an input and output ports on two given sites $m$ and $n$ as in Fig.~\ref{fig:sup_fig_2}. Each of these sites are connected to standard transmission lines with impedance $Z_0$ via two coupling capacitors $C_c$.
\begin{figure*}[ht]
	\centering
	\includegraphics[width = 1\textwidth]{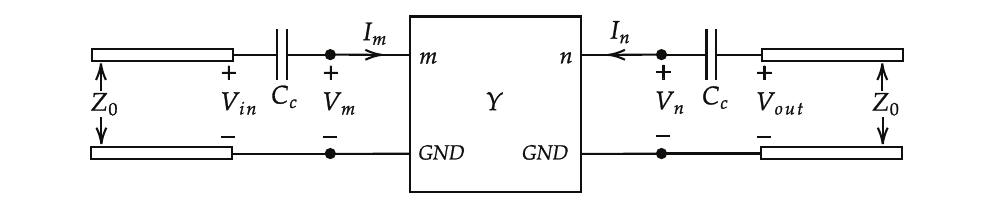}
    \caption{Finite lattice with admittance matrix $Y$, connected at two sites $m$ and $n$ to input and output transmission line using coupling capacitors $C_c$.}
    \label{fig:sup_fig_2}
\end{figure*}

The input voltage is divided to two forward and backward parts. The output voltage has only forward part: 
\begin{equation}
V_{\rm in} = V_{\rm in}^+ + V_{\rm in}^- ~,~ V_{\rm out} = V_{\rm out}^+. 
\label{v_in_out}
\end{equation}
Using transmission line theory \cite{SIpozar2011microwave} the currents on input (output) ports are related to input (output) voltages using $Z_0$ via
\begin{equation}
I_m = \frac{V_{\rm in}^+}{Z_0} - \frac{V_{\rm in}^-}{Z_0} ~,~ I_n = -\frac{V_{\rm out}^+}{Z_0}.
\label{I_in_out}
\end{equation}
Next we write Ohm's law for the left (right) coupling capacitor which has impedance $Z_c^{-1} = \ii\omega C_c$ as
\begin{equation}
V_{\rm in} - V_m = Z_c I_m ~,~ V_{\rm out} - V_n	= Z_c I_n.
\label{ohm_C_c}
\end{equation}
Our finite lattice with $N_s$ sites is defined with an admittance matrix which is obtained from KCL Eq. ~\eqref{kcl_eq} on each site. Here the external current exists only for sites $m$ or $n$ and the external current at all other sites is zero. The matrix equation for the finite lattice takes the form: 
\begin{equation*}
Y 
\begin{pmatrix}
	\vdots\\
	V_m\\
    \vdots\\
	V_n\\
	\vdots
\end{pmatrix} = 
\begin{pmatrix}
	\textbf{0}\\
	I_m\\
	\textbf{0}\\
	I_n\\
	\textbf{0}\\
\end{pmatrix}.
\end{equation*}
In this equation each \textbf{0} is vector whose length matches with the number of sites up to $V_m$, between $V_m$ and $V_n$ and below $V_n$. For simplicity we define $Y^{-1}=Z$. This is actually is the Green's function for our structure. By using Eq. \ref{YD_rel} we get the following relation for impedance matrix:
\begin{equation*}
    Z = \ii \omega L (D - \lambda \mathbb{I})^{-1}.
\end{equation*}
When we substitute $D$ by the eigenvalue matrix $\Gamma$ and eigenvector matrix $\tilde{V}$ the above relation is simplified to
\begin{equation}
    Z = \ii \omega L \tilde{V} (\Gamma - \lambda \mathbb{I})^{-1} \tilde{V}^{\dagger}
    \label{Z_eig_rel}.
\end{equation}
Since all external currents for the lattice except $m$'th and $n$'th sites are zero we get:
\begin{align*}
& V_m = Z_{mm}I_m + Z_{mn}I_n, \\
& V_n = Z_{nm}I_m + Z_{nn}I_n.	
\end{align*}
Note that using Eq. \ref{Z_eig_rel} we can find any element $Z_{mn}$ as follows:
\begin{equation*}
    Z_{mn}(\lambda) / (\ii \omega L ) = \sum_{jl} \tilde{V}_{mj}  \delta_{jl} \frac{1}{\lambda_j - \lambda}  \tilde{V}^{\dagger}_{ln} = \sum_j \tilde{V}_{mj} \tilde{V}^{*}_{nj} \frac{1}{\lambda_j - \lambda}.
\end{equation*}
Now we substitute $V_m$, $V_n$, $I_m$ and $I_n$ in terms of $V_{\rm in}^+$, $ V_{\rm in}^-$ and $V_{\rm out}^+$ using Eqs. \ref{v_in_out}, \ref{I_in_out} and \ref{ohm_C_c} to obtain:
\begin{align*}
V_{\rm in}^+ + V_{\rm in}^- - \frac{Z_c}{Z_0}(V_{\rm in}^+ - V_{\rm in}^-) &= Z_{mm} \frac{V_{\rm in}^+ - V_{\rm in}^-}{Z_0} + Z_{mn} \frac{-V_{\rm out}^+}{Z_0}, \\
V_{\rm out}^+(1 + \frac{Z_c}{Z_0}) &= Z_{nm} \frac{V_{\rm in}^+ - V_{\rm in}^-}{Z_0} + Z_{nn} \frac{-V_{\rm out}^+}{Z_0}.	
\end{align*}
By solving these two equations we obtain $V_{\rm in}^-$ and $V_{\rm out}^+$ in term of $V_{\rm in}^+$. Therefore the scattering parameters $S_{11}$ and $S_{21}$ which are the reflection and transmission from input and output ports will be given by:
\begin{align}
S_{11} & = \frac{(Z_{mm}+Z_c-Z_0)(Z_{nn}+Z_c+Z_0)-Z_{mn}Z_{nm}}{(Z_{mm}+Z_c+Z_0)(Z_{nn}+Z_c+Z_0)-Z_{mn}Z_{nm}}, \\
S_{21} & = \frac{2Z_0Z_{nm}}{(Z_{mm}+Z_c+Z_0)(Z_{nn}+Z_c+Z_0)-Z_{mn}Z_{nm}}.
\label{S21SIgeneral.eqn}
\end{align}
When we tune the excitation frequency near one of resonances (eigenvalues) for example $\lambda_j = (\omega_j/\omega_0)^2$ each of impedance matrix elements in $S_{21}$ relation is approximated by its leading term,
\begin{equation*}
    Z_{mn} \approx \tilde{V}_{mj} \tilde{V}^{*}_{nj} \frac{\ii \omega_j L}{\lambda_j - \lambda}.
\end{equation*}
Then we substitute these approximate impedance matrix elements in $S_{21}$ and simplify to obtain,
\begin{equation*}
    S_{21} \approx \frac{2 Z_0 \tilde{V}_{nj}\tilde{V}^*_{mj}}{(Z_c + Z_0) (|\tilde{V}_{nj}|^2 + |\tilde{V}_{mj}|^2)} \frac{1}{1 + \frac{(\lambda_j - \lambda)(Z_c + Z_0)}{\ii\omega_jL(|V_{mj}|^2+|V_{nj}|^2)}}.
\end{equation*}
When we substitute $\lambda_j$ and $\lambda = (\omega_j+\delta\omega)/\omega_0$ we get a Lorentzian behavior around each resonance. 

\section{Design and simulation}
\subsection{ Designing the value of lattice capacitor and inductance values}

\subsection{Finite lattice mode shapes}
To approach the frequency spectrum of a very large (effectively infinite) network and examine any distortions within it, we need to increase the network size. However, due to synthesis limitations, we can practically only increase the number of network sites to a certain extent. In this section, we proceed with the design for a network with 731 sites. 
\begin{figure}[t]
	\centering
	\includegraphics[width = 0.65\textwidth]{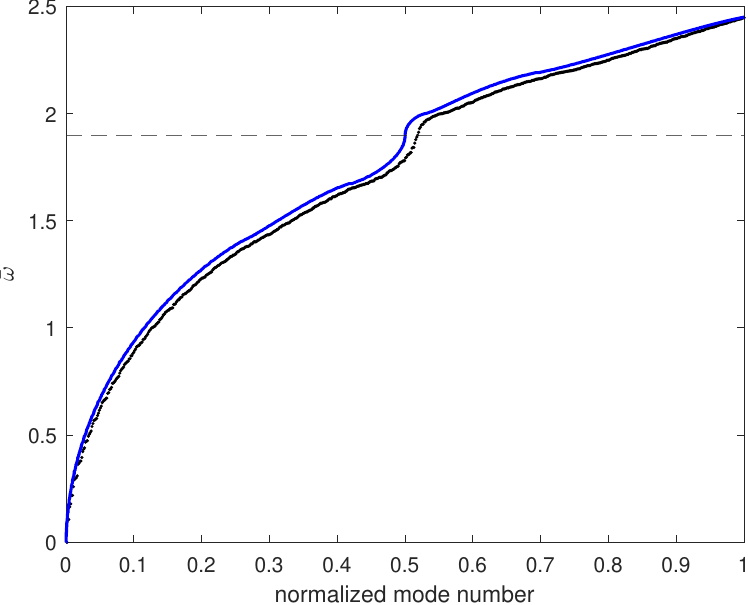}
	\caption{Resonance mode diagrams (black dots) for a network with limited dimensions and a specified number of sites based on mode number - for comparison, the resonance frequencies of an infinite network with translation symmetry (blue curve) are also plotted. The dashed line indicates the Dirac frequency. All frequencies are plotted in the normalized state. Since the number of modes depends on the number of sites, to align the horizontal axes of the two graphs for comparison, normalization is performed by dividing the mode number by the total number of modes.}
	\label{fig_fin_lattice_spect}
\end{figure}
Figure~\ref{fig_fin_lattice_spect} shows the connection design of this network. Additionally, in the graph of Figure~\ref{fig_fin_lattice_spect}, the frequency spectrum of this network is compared with that of an infinite network for the same value of $\zeta = 0.4$. As is evident, these two spectra are very close to each other.

\begin{figure}[t] \centering \includegraphics[width = 0.95\textwidth]{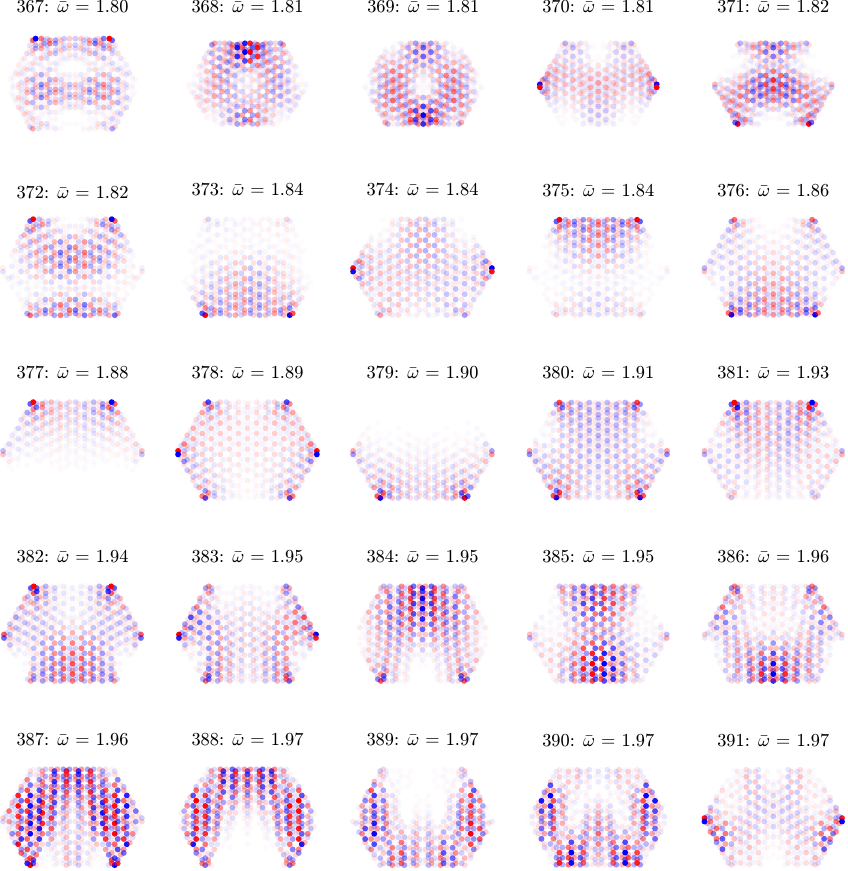} \caption{\textbf{ Intensity representation for the eigenvectors on the network.} The mode number and its normalized frequency are specified above each figure. Note that as the intensity increases, the points become darker. Also, red and blue colors indicate the positive or negative sign of $S_{2}$ \textcolor{red}{or $S_{21}$}.} \label{fig_fin_lattice_731_eigenVec} \end{figure}

Figure~\ref{fig_fin_lattice_731_eigenVec} shows the intensity of the eigenvectors on the network for 25 modes around the Dirac frequency. As observed, the intensity concentration in different modes varies across different sites. In modes where the intensity concentration is on the edges, the impact of the edge and open boundary condition is more pronounced, for example, see modes 373 to 381. By examining all the eigenvectors, it is generally observed in this structure that in the edge modes, the intensity is concentrated at the corners, and the intensity on the boundary edge is much lower. This insight helps us choose the location of the input and output signals to excite the network. Considering that we aim to excite the bulk resonance frequencies within the network as much as possible and reduce edge effects as much as possible,  we avoid the corners and connect the input/output port to the middle of the edges. 
\begin{figure*}[t]
	\centering
	\includegraphics[width = 1\textwidth]{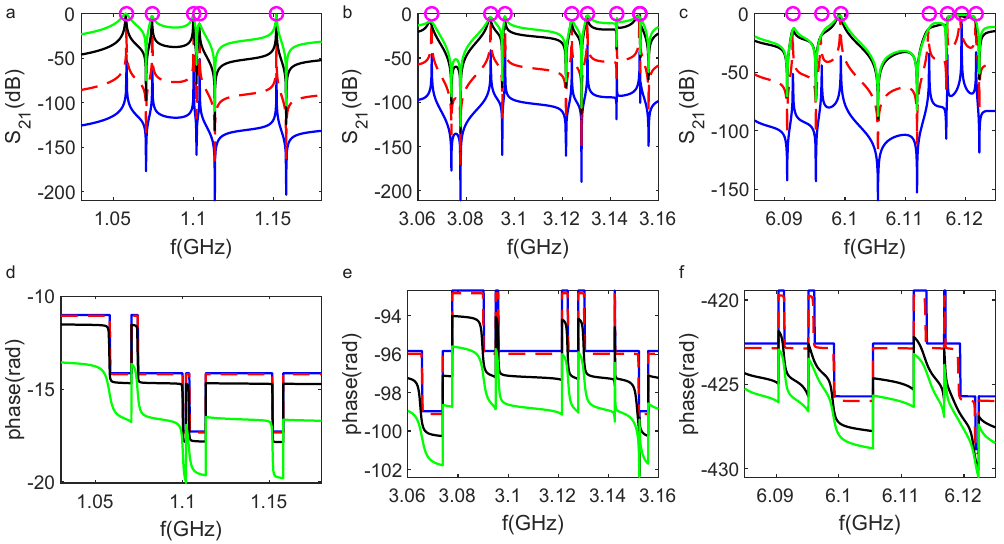}
    \caption{a, b and c show $S_{21}(dB)$ for three frequency ranges. The corresponding phase of $S_{21}$ for each range is shown below it. The blue, red, black and green colors are for $\varrho = 0.001,~0.01,~0.1,~1$ respectively.}
    \label{s21_coupling}
\end{figure*}

\subsection{Coupling Capacitor Optimum Value}
Using the  relation~\eqref{S21SIgeneral.eqn} for $S_{21}$ we can calculate the scattering matrix for a finite lattice for different values of coupling capacitor. We define $\varrho = C_c/C$ as the coupling ratio. Our Lattice has 731 sites and the parameters of lattice are: $C = 8.07~pF$, $L = 0.37~nH$ and $\zeta = \frac{2}{5.02} \approx 0.4$. Solving eigenvalue problem we first find the resonances (eigenvalues) and their corresponding eigenvectors. The input and output locations is on the edges of the lattice as discussed in the previous sub-section. In Fig. \ref{s21_coupling} we plot the magnitude and phase of $S_{21}$ on different range of frequencies for four coupling ratios $\varrho = 0.001,~0.01,~0.1,~1$. As can be seen, smaller $\varrho$ produces sharper peaks with lower line-width. Also the phase has higher slope at resonance. But the main problem is that the signal is very weak. For example for $\varrho = 0.001$ (blue curves) the peaks have values around $-50~dB$ which is very low for practical purposes. As $\varrho$ increases the line-width grows which hinders the peak detection process, see e.g. $\varrho = 1$ (green curves). The best range for $\varrho$ is between $0.01$ to $0.1$. We choose $\varrho = 0.1$ in order to compensate for parasitic effects in fabricated structure.



\section{Data processing}
\subsection{Peak Detection Algorithm}
In this section, we explain how to find resonances from the measured data of scattering matrix $S_{21}$. Normally, at each resonance, the magnitude ($|S_{21}|$) and phase gradient (PhG) of $S_{21}$ indicates a peak. Generally, we can look at PhG and find its resonances using peak detection algorithms. In such algorithms, first we filter PhG to make it smoother and attenuate noise effects using the Savitzky-Golay filtering method~\cite{SISavGolFlt}. Next, the frequency range is divided into small slices, and in each slice, the average and standard deviation (std) of PhG are calculated. Then, if the height of a PhG peak crosses the average more than one standard deviation, then the peak is counted. There are two sources for the error in peak locations: Due to unwanted noises in some frequency ranges the detection is hard and there can be fake peaks in noisy sections. Also in some frequencies the height of PhG peaks are very low and not able to cross the desired height level leading to missing some peaks.  Therefore, to make sure we detect the right resonances, we additionally use the magnitude peak to find resonances. In this method, often due to the Fano resonance effect \cite{SILukyanchuk2010}, the $|S_{21}|$ behavior near resonance has a peak-dip character. To identify such Fano resonances, we find local maxima and minima of $|S_{21}|$ and compare the distance between them. If a peak and a dip frequency distance is lower than a threshold we record them as one resonance. Also, we keep the remaining resonances corresponding to a peak or dip. Finally, when we determine the resonance frequencies using both magnitude and PhG we contrast them. If two resonances from both methods are close to each other (in the scale of their line-widths) we record it as one resonance. In the end, we add the remaining resonances of $|S_{21}|$ to our final list to compensate for the resonances missed by PhG method due to their low hight. 

In Figs.~\ref{fig:sup_fig_4} and~\ref{fig:sup_fig_5}, we show the results of the resonance detection method over the entire frequency range of measurement (0.5~GHz to 10~GHz) in 8 intervals with the range of 300~MHz for the structure with the designed value of $\zeta = \frac{2}{5.02} \approx 0.4$. In Figs. \ref{fig:sup_fig_4}a and \ref{fig:sup_fig_4}b, PhG is very noisy and hence we use $|S_{21}|$ for peak detection. Note that in this region there are only peaks in  $|S_{21}|$ with no dips. In other frequency ranges, besides PhG and $|S_{21}|$ peaks, the sharp dips of $|S_{21}|$ help us to locate the resonances more accurately. In the frequency range of Fig. \ref{fig:sup_fig_5}b, because we are around Dirac point, number of resonances significantly decreases. Above this range in Figs. \ref{fig:sup_fig_5}c and \ref{fig:sup_fig_5}d, again the number of resonances sharply increases. Also, the DOS in these regions is larger than in the intervals around the Dirac points.  

\begin{figure*}[t]
	\centering
	\includegraphics[width = 1\textwidth]{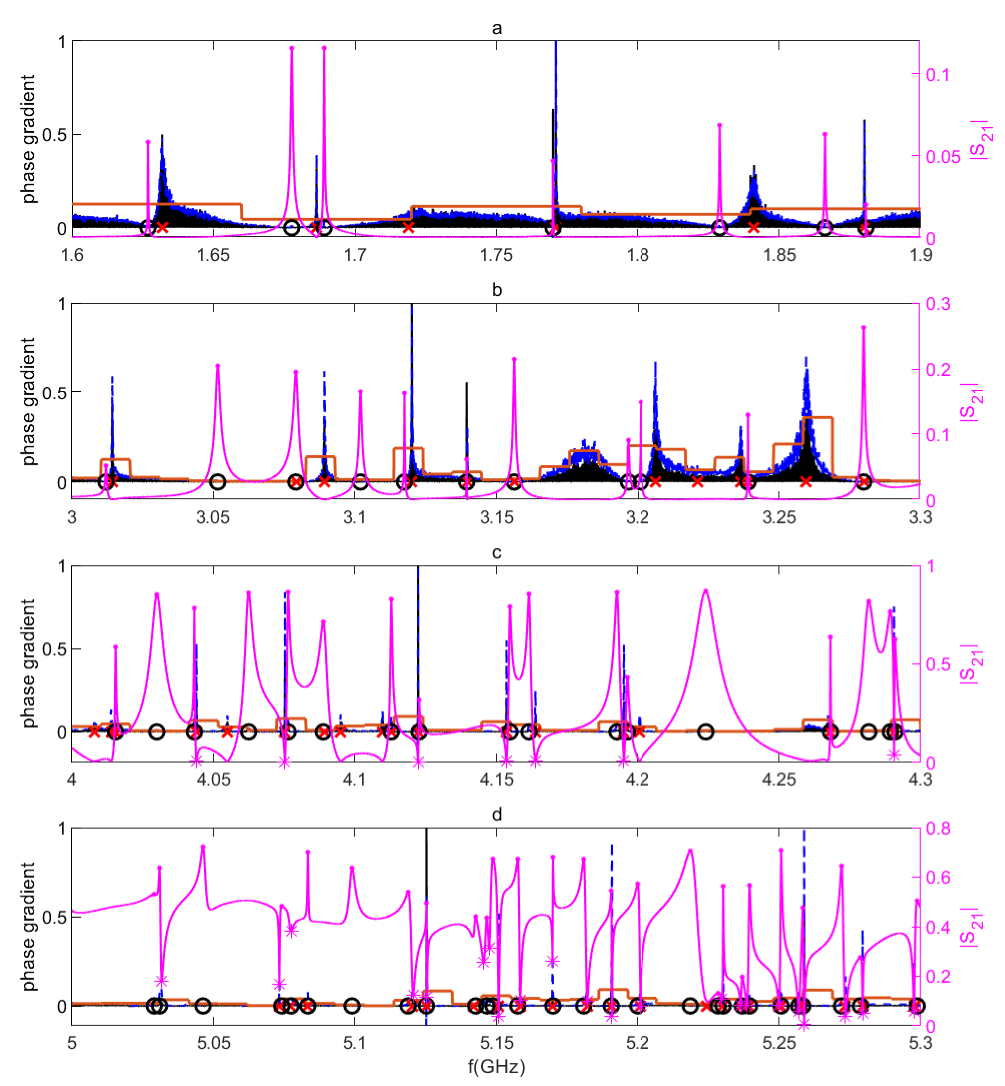}
    \caption{Detection of resonance frequencies for a selected frequency section. In all of these figures, we show PhG by black solid line, filtered PhG by dashed blue line, baseline for PhG peak detection by orange solid line, $|S_{21}|$ by magenta solid line, peaks of $|S_{21}|$ by magenta dots, dips of $|S_{21}|$ by asterisks, PhG peaks by crossed red points and selected peaks by black circle points. (a) frequency range 1.6-1.9~GHz. (b) frequency range 3.0-3.3~GHz. (c) frequency range 4.0-4.3~GHz. (d) frequency range 5.0-5.3~GHz.}
    \label{fig:sup_fig_4}
\end{figure*}
\begin{figure*}[t]
	\centering
	\includegraphics[width = 1\textwidth]{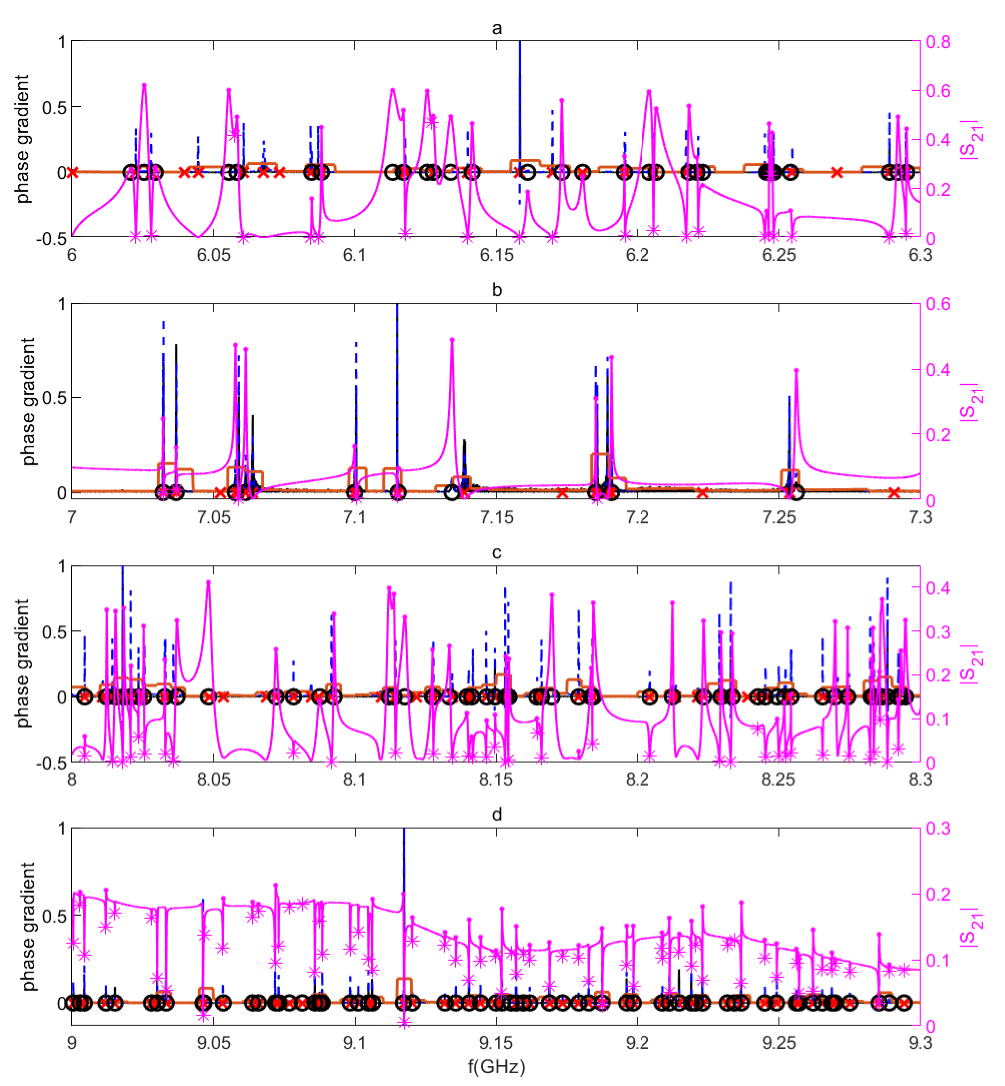}
    \caption{Same as previous figure: detection of resonance frequencies for another frequency section (a) frequency range 6.0-6.3~GHz. (b) frequency range 7.0-7.3~GHz. (c) frequency range 8.0-8.3~GHz. (d) frequency range 9.0-9.3~GHz.}
    \label{fig:sup_fig_5}
\end{figure*}

\subsection{Determination of the bandwidth of resonances}
After finding resonance frequencies we have to determine the bandwidth of each resonance in order to ensure the resonances are sharp enough with narrow bandwidth. We divide the whole frequency in segments each containing one resonance. The location of divisions are between two neighbouring resonances. After that for each segment we fit $|S_{21}|^2$ to a Lorentz oscillator model with Fano form~\cite{SILimonov2017}:
\begin{equation}
    |S_{21}|^2 = A\left[\frac{(\Omega+Q^{-1}q/2)^2}{\Omega^2+(Q^{-1}/2)^2}\eta + (1-\eta)\right].
\end{equation}
In this equation $\Omega = (f/f_0)^2-1$ is the normalized frequency, $Q = f_0/\kappa$ is the quality factor where $\kappa = \Delta f$ is the line-width of each resonance. The $0\leq\eta\leq1$ along with the Fano-parameter $q$ (see below) contains the effect of interference with other resonances on scattering parameters near this resonance. In Fig.~\ref{fig:sup_fig_6} the fitting curve over $|S_{21}|^2$ for $6$ resonances are demonstrated. In each figure the value of Lorentzian model parameters are shown on top of the figure. As can be seen in Figs.~\ref{fig:sup_fig_6}a and~\ref{fig:sup_fig_6}b , for large values $|q|$ the Fano effect is negligible and the normal shape of resonance with a sharp peak is obtained. In Figs.~\ref{fig:sup_fig_6}e and~\ref{fig:sup_fig_6}f, the value of $q$ is near zero, therefore at resonance frequency there is sharp dip. When $q$ is around $\pm 1$, the curve near resonance has both peak and dip behavior, which is seen in Figs.~\ref{fig:sup_fig_6}c and~\ref{fig:sup_fig_6}d. In all of these figures, the model is fitted very well near the resonance. It is obvious that by getting away from resonance because of the presence of neighboring resonances the $|S_{21}|^2$ curve is not fitted to the model anymore. Sometimes resonance may happen on the ascending or descending slope of another resonance, so in our fitting process it is better to add a linear term to our model. This term has negligible effect on the value of other parameters.

\begin{figure*}[t]
	\centering
	\includegraphics[width = 1\textwidth]{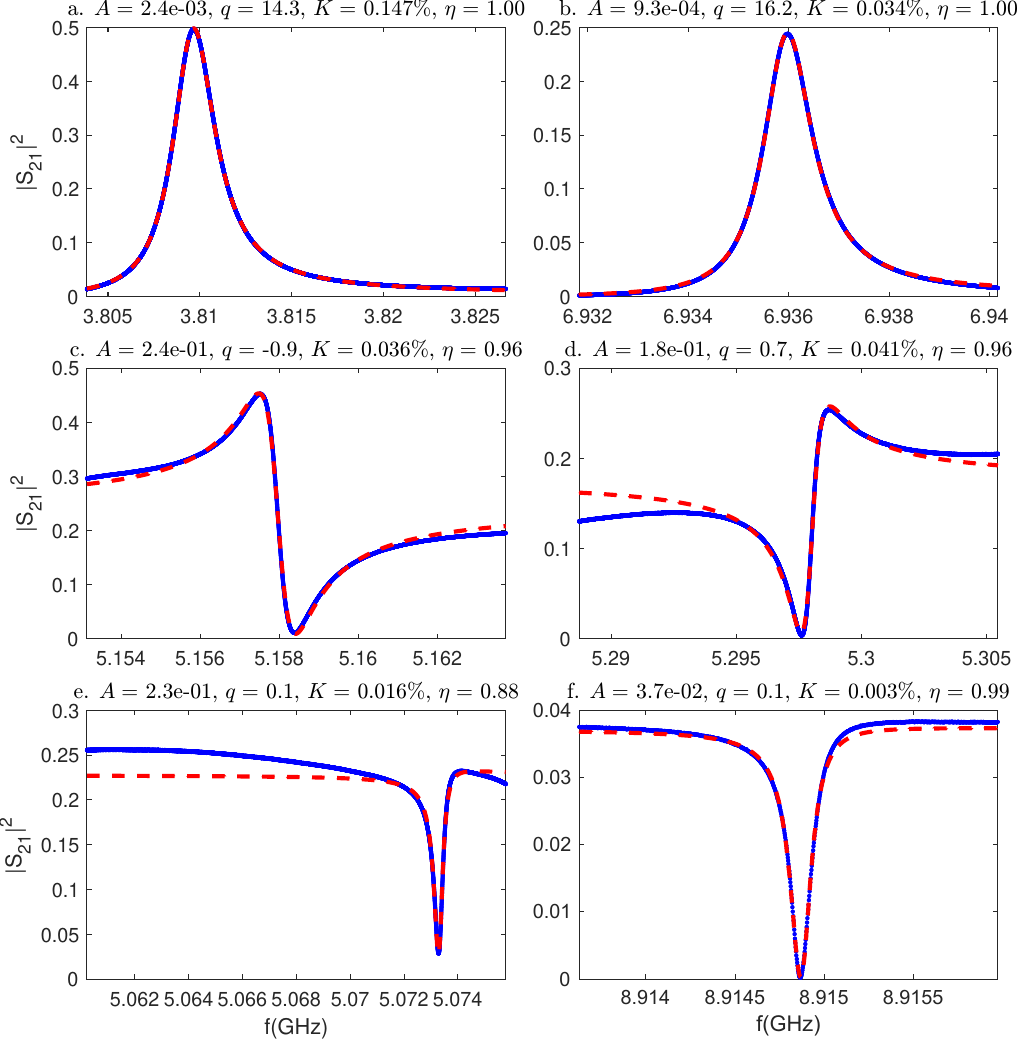}
    \caption{Fitting Lorentzian model (red dashed line) to $|S_{21}|^2$ (blue line) for six different resonances. (a) and (b): In these resonances Fano effect is weak and we see sharp peak. (c) and (d): Peak and dip behavior of $|S_{21}|^2$ around resonance. (e) and (f): In these resonances $q$ is near zero, so we have sharp dips.}
    \label{fig:sup_fig_6}
\end{figure*}

Now, after finding the line-width of each resonance, we present their histogram in Fig.~\ref{fig:sup_kappa_hist}.
As can be seen, around 62\% of resonances have line-width less than 1~MHz and around 85\% have line-width less than 2~MHz. 
\begin{figure*}[t]
	\centering
	\includegraphics[width = 0.8\textwidth]{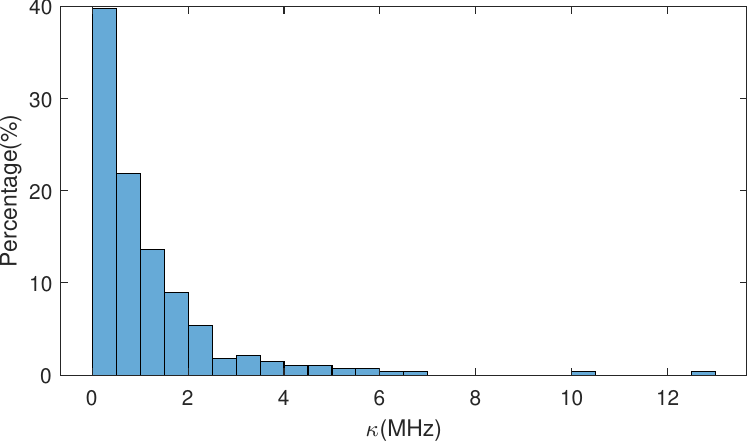}
    \caption{Line-width histogram- The histogram counts the number of line-width in each bin with frequency length of 0.5~MHz. In this plot only the line-width of resonances around Dirac frequency, which is 7.3~MHz, in the range 6~MHz to 8.5~MHz are counted. The vertical axis shows the percentage of counted number in each bin to the total counts.}
    \label{fig:sup_kappa_hist}
\end{figure*}

In the Fig.~\ref{fig:sup_quality_hist} the quality factors of all the detected resonances are shown. As can be seen nearly all of the quality factors are above 1000.
\begin{figure*}
	\centering
	\includegraphics[width = 0.8\textwidth]{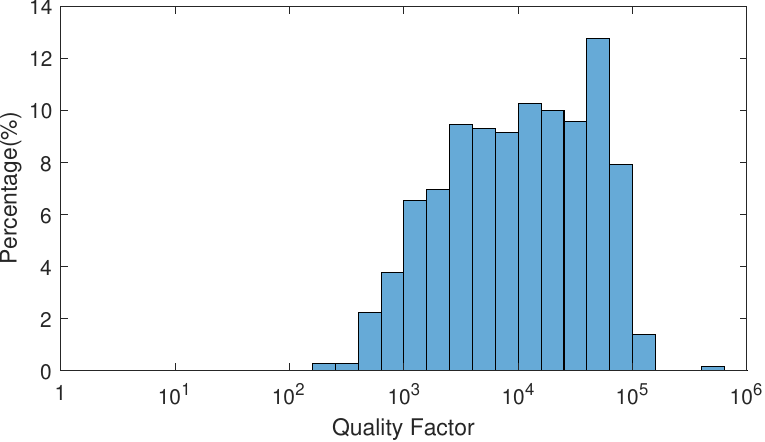}
    \caption{Quality factor histogram}
    \label{fig:sup_quality_hist}
\end{figure*}
\subsection{Fitting the Model to Detected Peaks}
In this subsection we explain how to find the parameters of the lattice in a way that its resonances fit to the detected resonances as close as possible. The finite lattice has 731 sites and therefore 731 resonances. Due to noise and overlap of some of the resonances the number of detected peaks are slightly lower than the number of sites in the lattice. In other words our peak detection algorithm always misses some of the resonances, due to noise effects. For example, for initially designed value of $\zeta = \frac{2}{5.02} \approx 0.2$, we detected 720 resonances. Therefore, we need a starting point in the whole frequency range to begin fitting the measured and lattice resonance frequencies. In frequencies less than 1~GHz we have a lot of noise, so it is possible that some of the missed frequencies are buried in this frequency range. Also, in the other end of measurement range around 9~GHz, resonances are very close to each other and their peaking/dipping behavior on $|S_{21}|$ or peaks in PhG are very hard to detect. Finally, around Dirac point, we know the number of resonances substantially decrease compared to other frequency regions. Hence, the best starting point for fitting is Dirac resonance frequency itself. We have 2 parameters to find: $\omega_0$ and $\zeta$. The initial guess values of these parameters are their designed values. Now using Eq.~\ref{eq: Dirac_frquency} we find the Dirac frequency as the starting point of the fitting procedure. The resonance frequency which is the closest to this frequency in both measured and model resonances labeled the same resonance. All other resonances of measured data and the model are paired to each other relative to this point. After this, We use the standard least square fitting algorithm to minimize distance between measured and the model resonance frequencies which is defined as follows by optimizing the value of parameters in each iteration:
\begin{equation}
    d = \sum_{i} (f^{({\rm measured})}_i - f^{({\rm model})}_i)^2,
\end{equation}
Where $i$ runs over the number of measured resonances. 

In Fig.~\ref{fig:sup_fig_8} the fitting of resonance modes for both measured data and the model are shown for each of the three designed value of $\zeta/2 = 0,~0.2,~0.25$. The designed value of $f_0$ for all three structures based on $C$ and $L$ is 2.91~GHz. As can be seen in all panels of this figure, model resonances and measured resonances are in excellent agreement. The fitting parameters of each structure is shown on top of each figure. Note that the error is obtained using least square algorithm. The value of $f_0$ is nearly the same for all structures as we expected as the value of $f_0$ is dependent only on $L$ and $C$ which are the same in all structures. Although, the designed value of $f_0$ differs from its fitted value, but the important thing is that all three structures have same values. The value of $\zeta$ in each structure is close to designed value which shows that the ratio between $L'$ and $L$ is less sensitive to possible errors in the entire procedure including design, fabrication and measurements.
\begin{figure*}[t]
	\centering
	\includegraphics[width = 0.8\textwidth]{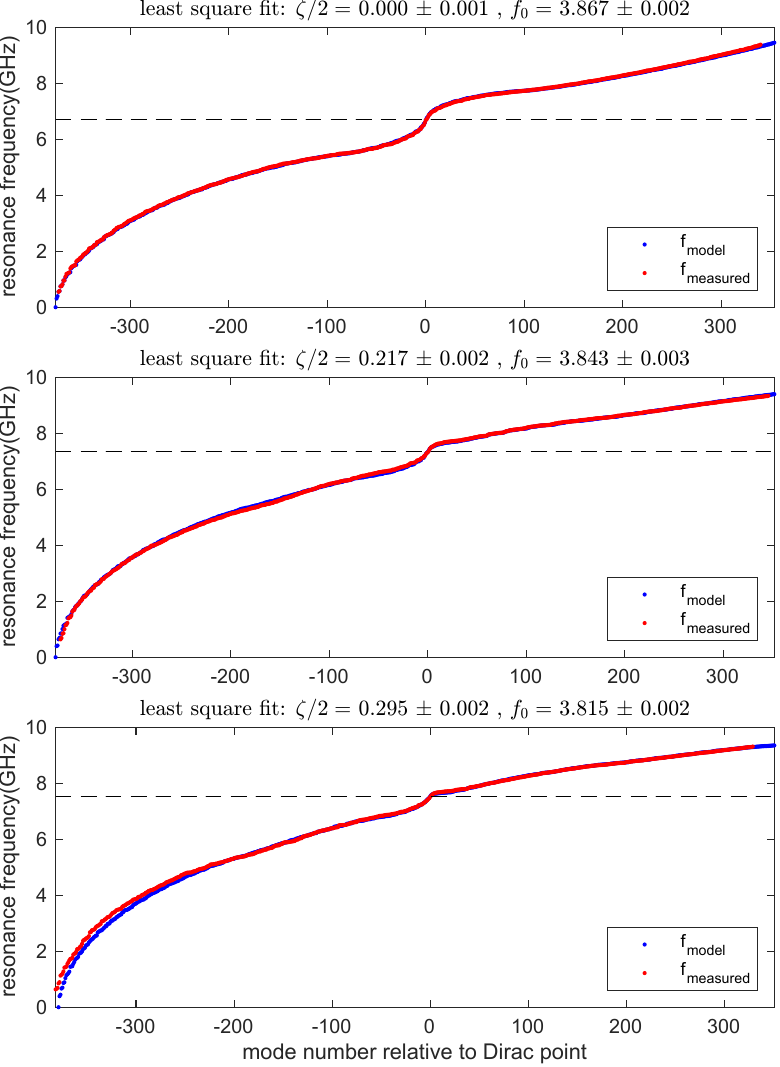}
    \caption{Comparison of measured (red) and fitted model (blue) resonances for each device. The horizontal line indicates the Dirac frequency for the corresponding device. The negative (positive) "mode numbers" relative to Dirac points represents the lower (higher) resonance frequencies.}
    \label{fig:sup_fig_8}
\end{figure*}
\newpage
\bigskip
\section{Supplementary References}

\bigskip

\putbib[Refs]
\end{widetext}
\end{bibunit}
\end{document}